\documentclass[twocolumn,aps,pr,superscriptaddress,preprintnumbers,nofootinbib,10pt]{revtex4-2}
\usepackage{multirow}
\usepackage{amsmath}
\usepackage{amssymb}
\usepackage[dvipdf,dvips]{graphicx}
\usepackage{color}
\usepackage{hyperref}
\usepackage{url}
\usepackage{slashed}
\usepackage{subfigure}
\usepackage[usenames,dvipsnames]{xcolor}
\usepackage{amsmath}
\usepackage{amsfonts}
\usepackage{float} 
\usepackage{amssymb}
\usepackage{epsfig}
\usepackage{graphics}
\usepackage{euscript}
\usepackage{slashed}
\usepackage{epstopdf}
\usepackage[utf8]{inputenc}
\allowdisplaybreaks
\usepackage[normalem]{ulem}
\usepackage{pifont}
\usepackage{dsfont}
\usepackage{MnSymbol}
\usepackage{verbatim}
\usepackage{graphicx}
\usepackage{latexsym}
\usepackage{tikz-feynman}
\usepackage{bbold}

\begin{document}

	\title{{\bf Unruh-De Witt detectors, Bell-CHSH inequality and Tomita-Takesaki theory}}

	
	\author{F. M. Guedes} \email{fmqguedes@gmail.com} \affiliation{UERJ $–$ Universidade do Estado do Rio de Janeiro,	Instituto de Física $–$ Departamento de Física Teórica $–$ Rua São Francisco Xavier 524, 20550-013, Maracanã, Rio de Janeiro, Brazil}
	
	\author{M. S.  Guimaraes}\email{msguimaraes@uerj.br} \affiliation{UERJ $–$ Universidade do Estado do Rio de Janeiro,	Instituto de Física $–$ Departamento de Física Teórica $–$ Rua São Francisco Xavier 524, 20550-013, Maracanã, Rio de Janeiro, Brazil}
	
	\author{I. Roditi} \email{roditi@cbpf.br} \affiliation{CBPF $-$ Centro Brasileiro de Pesquisas Físicas, Rua Dr. Xavier Sigaud 150, 22290-180, Rio de Janeiro, Brazil}
	
	\author{S. P. Sorella} \email{silvio.sorella@gmail.com} \affiliation{UERJ $–$ Universidade do Estado do Rio de Janeiro,	Instituto de Física $–$ Departamento de Física Teórica $–$ Rua São Francisco Xavier 524, 20550-013, Maracanã, Rio de Janeiro, Brazil}

	\begin{abstract}

The interaction between Unruh-De Witt spin $1/2$ detectors and a real scalar field is scrutinized  by making use of the Tomita-Takesaki modular theory as applied to the Von Neumann algebra of the Weyl operators. The use of the modular theory enables to evaluate in an exact way the trace over the quantum field degrees of freedom. The resulting density matrix is employed to the study of the Bell-CHSH correlator. It turns out that, as a consequence of the interaction with the quantum field,  the violation of the Bell-CHSH inequality exhibits a decreasing as compared to the case in which the scalar field is absent.

	\end{abstract}

	\maketitle

\section{Introduction}

The so-called Unruh-De Witt detectors serve as highly useful models  that are largely employed in the study of relativistic quantum information, see  \cite{Reznik:2002fz,Tjoa:2022vnq,Tjoa:2022lel} and refs. therein. \\\\In the  current work, we shall   utilize spin $1/2$  Unruh-De Witt detectors to  investigate the potential impact of a quantum relativistic scalar field on the Bell-CHSH inequality \cite{Bell64, CHSH69}. More precisely, we shall start by considering the interaction of a pair of q-bits with a real Klein-Gordon field in Minkowski spacetime. The initial state of the Klein-Gordon field is identified as  the vacuum state $| 0 \rangle$. Concerning the q-bits, the corresponding state will be taken as 
\begin{align}
    \vert \psi \rangle_{AB} = \frac{\vert g_{A} \rangle \otimes \vert g_{B} \rangle + r \vert e_{A} \rangle \otimes \vert e_{B} \rangle}{\sqrt{1+r^2}}, \; \; \; r \in [0,\, 1]\;, \label{stateqbit}
\end{align}
where, using the same notation of  \cite{Tjoa:2022vnq,Tjoa:2022lel},  $\vert g_{j} \rangle$, $\vert e_{j} \rangle$, $j=A,B$ stand for the ground and excited states  of the  two-level  Hamiltonian
\begin{align}
    h_{j} = \frac{1}{2} \Omega_{j} (\sigma_{j}^{z} + \mathbb{1}), \; \; \; \; j= A,\, B,
    \label{UDW free hamiltonian}
\end{align}
with $\sigma_{j}^{z}$ being the diagonal Pauli matrix along the $z$-direction.  The states  $\vert g_{j} \rangle$, $\vert e_{j} \rangle$  possess energy $0$ and $\Omega_{j}$, respectively. As it is customary in the study of entanglement, the indices $A,B$ refer to Alice and Bob which, according to the relativistic causality requirement, are located in the right and left Rindler wedges. Moreover, the parameter $r$ in expression \eqref{stateqbit} will enable us to interpolate between a product state, corresponding to $r=0$, and a maximally entangled state, {\it i.e.} when $r=1$. \\\\For the Hamiltonian describing the interaction between the q-bits and the real scalar field $\varphi(x)$, we have \cite{Tjoa:2022vnq,Tjoa:2022lel},
\begin{align}
    h_{Ij}(x) = f_{j}(x) \mu_{j}(\tau_{j}(x)) \otimes \varphi (x),
    \label{Interacting hamiltonian}
\end{align}
where
\begin{align}
\mu_{j}(\tau_{j})=\sigma_{j}^{+} e^{i \Omega_{j}\tau{j}} + \sigma_{j}^{-} e^{-i \Omega_{j}\tau{j}}
\label{monopole moment}
\end{align}
is the monopole moment of the detector $j$ with proper time $\tau_{j}$ \cite{Tjoa:2022vnq,Tjoa:2022lel}. The matrices $\sigma^{\pm}$ stand for the ladder operators
\begin{align*}
    \sigma_{j}^{+} \vert g_{j} \rangle &= \vert e_{j} \rangle \\
    \sigma_{j}^{-} \vert e_{j} \rangle &= \vert g_{j} \rangle.
\end{align*}
The functions $f_j(x)$ are smooth test functions with compact support, $f_j(x) \in  {\cal C}_{0}^{\infty}(\mathbb{R}^4)$. \\\\At this stage, we  need to specify the starting density matrix, namely 
\begin{align}
    \rho_{AB\varphi}(0) = \rho_{AB}(0)  \otimes  |0 \rangle    \langle 0| \;, \label{ddo}
\end{align}
where 
\begin{equation} 
\rho_{AB}(0)= \vert \psi  \rangle_{AB}\; _{AB} \langle \psi | \;. \label{qbrho}
\end{equation}
Furthermore, the time evolution of the initial density matrix, Eq.\eqref{ddo}, is  governed by the unitary operator 
\begin{align}
    \mathcal{U} = T e^{-i \int d^4x (h_{IA}(x) + h_{IB}(x))},
    \label{Unit operator}
\end{align}
where $T$ is the time ordering operator. For the density matrix at the very
large time, one has
\begin{align}
   \rho_{AB\varphi}=   \rho_{AB\varphi}(t\rightarrow \infty ) = \mathcal{U} \; \rho_{AB \varphi}(0)  \; \mathcal{U}^{\dagger} \;. \label{ddt}
\end{align}
The next step is that of  obtaining the density matrix ${\hat \rho}_{AB}$ for the q-bits system by tracing out the field modes: 
\begin{equation} 
{\hat \rho}_{AB} = {\rm Tr}_{\varphi} (\rho_{AB\varphi}) \;. \label{trace}
\end{equation} 
Finally, one is ready to evaluate the Bell-CHSH correlator 
\begin{eqnarray}  
\langle \mathcal{C} \rangle & = & {\rm Tr}({\hat \rho}_{AB} \mathcal{C})  \nonumber \\
\mathcal{C} & = & (A+A')B + (A-A')B' \;, \label{Cc}
\end{eqnarray}
where $(A,A')$, $(B,B')$ are the Bell operators, see Section \eqref{BCHSH}. In that way we are able to investigate the violation of the Bell-CHSH inequality by taking into account the effects arising from the presence of the quantum field $\varphi(x)$, encoded in the density matrix ${\hat \rho}_{AB}$. \\\\Having outlined the working setup, we proceed by stating our main result as well as by presenting the organization of the present work: 
\begin{itemize} 
\item the first aspect which we would like to  highlight is the role which will be played by the unitary Weyl operators 
\begin{equation} 
W_{f_j} = e^{i {\varphi}(f_j) } \;, \qquad j=A,B \;, 
\label{Weyli}
\end{equation} 
where $\varphi(f_j)$ is the smeared field \cite{Haag92}
\begin{equation} 
\varphi(f_j) = \int d^4x \; f_j(x) \varphi(x) \;. \label{smf}
\end{equation}
As one can figure out, these operators arise from the evolution operator $\mathcal{U}$, as discussed in Sect.\eqref{TTW}. It is well established that the operators $W(f_j)$ enjoy a rich algebraic structure, giving rise to a von Neumann algebra \cite{Haag92,Witten18,SW1,Summers87b,Weyl23}. In particular, from the Reeh-Schlieder theorem \cite{Haag92,Witten18}, it follows that the vacuum state $|0\rangle$ is both cyclic and separating for the aforementioned von Neumann algebra, 

\item These properties enable us to make use of the powerful Tomita-Takesaki modular theory \cite{Bratteli97}. As shown in \cite{SW1,Summers87b,Weyl23}, the  modular theory is very well suited for the algebra of the Weyl operators. In particular, as it will be discussed in Section \eqref{TTW}, the modular operators $(j,\delta)$ provide an exact evaluation of the correlation functions of the Weyl operators in terms of the inner products between Alice's and Bob's test functions $f_A$ and $f_B$. 

\item  As a consequence, as detailed in Section \eqref{BCHSH} and Section \eqref{dephasing}, the impact of the quantum field $\varphi$ on the violation of the Bell-CHSH inequality can be evaluated in closed form. Notably, it turns out that the violation of the Bell-CHSH inequality exhibits a decreasing behavior as compared to the case in which the field $\varphi$ is absent. This behavior is clearly visible through the exponential  factors arising from the correlation functions of the Weyl operators, as exemplified in equation \eqref{chsh dephasing}.

\end{itemize}

\section{Evaluation of the q-bits density matrix in the case of the $\delta$-coupled detectors}

Let us begin the study of the denisty matrix ${\hat \rho}_{AB}$ by considering the so-called $\delta$-coupling  \cite{Tjoa:2022vnq,Tjoa:2022lel}, corresponding to the regime in which the interaction between the q-bits and the scalar field $\varphi$ occurs at very short timescales, described thorugh $\delta$-functions of the proper times of the two detectors $(A,B)$. Following  \cite{Tjoa:2022vnq,Tjoa:2022lel}, the evolution  operator is given by $\mathcal{U} = \mathcal{U}_{A} \otimes \mathcal{U}_{B}$, where the unitary operator for the detector $j=A,B$ is
\begin{align}
    \mathcal{U}_j = e^{-i \mu_{j}(\tau_{j0}) \otimes \varphi(f_{j})},
    \label{Unitary op j}
\end{align}
with the commutation relation
\begin{align}
\left [ \mathcal{U}_{A}, \mathcal{U}_B \right ] = 0.
\end{align}
Using the algebra of the Pauli matrices, it is easy to show that expression \eqref{Unitary op j} can be written as 
\begin{align}
    \mathcal{U}_j = \mathbb{1} \otimes c_{j} - i \mu_{j}(\tau_{j0}) \otimes s_{j},
\end{align}
where $c_{j} \equiv \cos \varphi (f_{j})$ and $s_{j} \equiv \sin \varphi (f_{j})$. Given the initial matrix density $\rho_{AB\varphi}(0)$, Eq.\eqref{ddo}, its evolution reads 
\begin{align}
    \rho_{AB \varphi} &= \left ( \mathcal{U}_{A} \otimes \mathcal{U}_{B} \right ) \rho_{AB\varphi}(0) \;\mathcal{U}_{A}^{\dagger} \otimes \mathcal{U}_{B}^{\dagger} \nonumber \\
    &= \left( \mathbb{1}_{A} \otimes \mathbb{1}_{B} c_{A} c_{B} \right. \nonumber \\
    &\quad - i \mathbb{1}_{A} \otimes \mu_{B} c_{A} s_{A} \nonumber \\
    &\quad - i \mu_{A} \otimes \mathbb{1}_{B} s_{A} c_{B} \nonumber \\
    &\quad \left. - \mu_{A} \otimes \mu_{B} s_{A} s_{B}\right)
     \rho_{AB}(0)  
    \otimes \vert 0 \rangle \langle 0 \vert \nonumber \\
    &\times \left( \mathbb{1}_{A} \otimes \mathbb{1}_{B} c_{A} c_{B} \right. \nonumber \\
    &\quad + i \mathbb{1}_{A} \otimes \mu_{B} c_{A} s_{A} \nonumber \\
    &\quad + i \mu_{A} \otimes \mathbb{1}_{B} s_{A} c_{B} \nonumber \\
    &\quad \left. - \mu_{A} \otimes \mu_{B} s_{A} s_{B} \right)  \;.
    \label{rhoABphi delta}
\end{align}
Taking the trace over $\varphi$, we get
\begin{align}
    \hat{\rho}_{AB} = & \rho_{AB}(0) \langle c_{A}^{2} c_{B}^{2} \rangle - \rho_{AB}(0) (\mu_{A} \otimes \mu_{B}) \langle c_{A}c_{B}s_{A}s_{B} \rangle \nonumber \\
    & + (\mathbb{1}_{A} \otimes \mu_{B}) \rho_{AB}(0) (\mathbb{1}_{A} \otimes \mu_{B}) \langle c_{A}^{2}s_{B}^{2} \rangle \nonumber \\
    & + (\mathbb{1}_{A} \otimes \mu_{B}) \rho_{AB}(0) (\mu_{A} \otimes \mathbb{1}_{B}) \langle c_{A}s_{A} c_{B}s_{B} \rangle \nonumber \\
    & + (\mu_{A} \otimes \mathbb{1}_{B}) \rho_{AB}(0) (\mu_{A} \otimes \mathbb{1}_{B}) \langle c_{A}c_{B}s_{A}s_{B}\rangle \nonumber \\
    & + (\mu_{A} \otimes \mathbb{1}_{B}) \rho_{AB}(0) (\mu_{A} \otimes \mathbb{1}_{B}) \langle s_{A}^{2} c_{B}^{2} \rangle \nonumber \\
    & - (\mu_{A} \otimes \mu_{B}) \rho_{AB}(0) \langle c_{A}s_{A} c_{B}s_{B} \rangle \nonumber \\
    & + (\mu_{A} \otimes \mu_{B}) \rho_{AB}(0) (\mu_{A} \otimes \mu_{B}) \langle s_{A}^{2} s_{B}^{2} \rangle \;,
    \label{trace rhoABphi over phi delta}
\end{align}
where $\langle c_A c_B s_A s_B\rangle$, etc., denotes the expectation value of the Weyl operators, namely 
\begin{equation}
\langle c_A c_B s_A s_B\rangle = \langle 0| c_A c_B s_A s_B |0\rangle \label{cvphi} \;. 
\end{equation}
As we shall see in the next Section, these correlation functions will be handled in closed form by means of the Tomita-Takesaki theory. Once the density matrix ${\hat \rho}_{AB}$ is known, one can proceed with the investigation of the Bell-CHSH correlator, {\it i.e.}
\begin{eqnarray} 
\langle {\cal C} \rangle & = & {\rm Tr}( {\hat \rho}_{AB }\; {\cal C} )   \nonumber  \\
{\cal C} & = & (A-A')B + (A-A')B' \;, \label{ccorr}
\end{eqnarray}
where $(A,A')$ and $(B,B')$ stand for Alice's and Bob's Bell's operators: 
\begin{eqnarray} 
A & = & A^{\dagger} \;, \quad A'= A'^{\dagger}\;, \quad B= B^{\dagger} \;, \quad B'= B'^{\dagger} 
\nonumber \\
A^2 & = & A'^2 = B^2=B'^2=1  \nonumber \\
\left[A,B \right] & = & [A,B'] = [A',B] = [A', B'] =0  \;. \label{Bop}
\end{eqnarray}  
The Bell-CHSH inequality is said to be violated whenever 
\begin{equation} 
2 < | \langle {\cal C} \rangle | \le 2 \sqrt{2} \;, \label{tsi}
\end{equation}
where the maximum value $2\sqrt{2}$ is known as the Tsirelson bound \cite{TSI}. The detailed analysis of Eq.\eqref{ccorr} can be found in Section \eqref{BCHSH}.

\section{Tomita-Takesaki modular theory  theory and the von Neumann algebra of the Weyl Operators}\label{TTW}

In order to face the evaluation of the correlation functions of the Weyl operators, Eq.\eqref{cvphi}, it is worth to provide a short account on some basic features of the properties of the related von Neumann algebra\footnote{See ref.\cite{Weyl23} for a more detailed account.}. Let us begin by reminding the expression of the causal Pauli-Jordan distribution $\Delta_{PJ}(x-y)$:
\begin{eqnarray}
[\varphi(x), \varphi(y)] & = & i \Delta_{PJ}(x-y) \nonumber \\
    i \Delta_{PJ}(x-y) \! &=&\!\! \int \!\! \frac{d^4k}{(2\pi)^3} \varepsilon(k^0) \delta(k^2-m^2) e^{-ik(x-y)} \;, \nonumber \\
    \label{PJ}
\end{eqnarray}
with $\varepsilon(x) \equiv \theta(x) - \theta(-x)$, As it is well known, $\Delta_{PJ}(x-y)$ is Lorentz invariant and vanishes when $x$ and $y$ are space-like
\begin{equation} 
\Delta_{PJ}(x-y) = 0 \;, \quad {\rm for} \quad (x-y)^2<0 \;. \label{spl}
\end{equation}
Let $\mathcal{O}$ be and open region of the Minkowski spacetime and let $\mathcal{M}(\mathcal{O})$ be the space of test functions $\in \mathcal{C}_{0}^{\infty}(\mathbb{R}^4)$ with support contained in $\mathcal{O}$:
\begin{align} 
	\mathcal{M}(\mathcal{O}) = \{ f \, \vert supp(f) \subseteq \mathcal{O} \}. \label{MO}
\end{align}
One introduces the symplectic complement \cite{SW1,Summers87b} of $\mathcal{M}(\mathcal{O})$ as 
\begin{align} 
	\mathcal{M'}(\mathcal{O}) = \{ g \, \vert  \Delta_{PJ}(g,f) = 0, \; \forall f \in \mathcal{M}(\mathcal{O}) \}, \label{MpO}
\end{align}
that is, $\mathcal{M}(\mathcal{O})$ is given by the set of all test functions for which the smeared Pauli-Jordan expression $\Delta_{PJ}(f,g)$  vanishes for any $f$ belonging to $\mathcal{M}(\mathcal{O})$ 
\begin{equation} 
\left[ \varphi(f), \varphi(g) \right] =  i  \Delta_{PJ}(f,g) \;. \label{smpj}
\end{equation} 
The symplectic complement $\mathcal{M'}(\mathcal{O})$ allows us to rephrase causality, Eq.\eqref{spl}, as \cite{SW1,Summers87b}
\begin{align}
    \left[ \varphi(f), \varphi(g) \right] = 0,
\end{align}
whenever $f \in \mathcal {M}(\mathcal{O})$ and $g \in \mathcal {M'}(\mathcal{O})$. \\\\We proceed by introducing the Weyl operators \cite{SW1,Summers87b,Weyl23}, a class of unitary operators obtained by exponentiating the smeared field 
\begin{equation} 
W_{h} = e^{i {\varphi}(h) }. 
\label{Weyl}
\end{equation}
Using the Baker–Campbell–Hausdorff formula and the commutation relation \eqref{PJ}, it turns out  that the Weyl operators give rise to the following algebraic structure:
\begin{align} \label{algebra} 
	W_{f} W_{g}   &= e^{ - \frac{i}{2} \Delta_{\textrm{PJ}}(f, g)}\; W_{(f+g)}, \nonumber \\
	W_{f}^{\dagger} W_{f} &= 	W_{f} W_{f}^{\dagger} = 1, \nonumber \\ 
	W^{\dagger}_{f} &=  W_{(-f)}.	 
\end{align} 
Furthermore, for $f$ and $g$ space-like, the Weyl operators $W_{f}$ and $W_{g}$ commute. Expanding the field in terms of creation and annihilation operators, see \cite{Weyl23},   one can compute the expectation value of the Weyl operator, finding
\begin{equation} 
\langle 0 \vert  W_{h}  \vert 0 \rangle = \; e^{-\frac{1}{2} {\lVert h\rVert}^2}, 
\label{valueW}
\end{equation} 
where $\vert\vert h \vert\vert^2 = \langle h \vert h \rangle$ and 
\begin{eqnarray} 
\langle f \vert g \rangle & = & \int \frac{d^3k}{(2\pi)^3} \frac{1}{2\omega_k} f(\omega_k,\vec{k})^{*} g(\omega_k,\vec{k}) \nonumber \\
f(\omega_k, \vec{k}) &=& \int d^4x \; e^{ikx} f(x) \;,  \label{inner}
\end{eqnarray} 
is the Lorentz invariant inner product between the test functions $(f,g)$\footnote{For $\omega_k$ we have the usual relation $\omega^2_k = k^2+m^2$. } \cite{SW1,Summers87b,Weyl23}. Taking now all possible products and linear combinations of the Weyl operators defined on $\mathcal{M}(\mathcal{O})$, gives rise to a von Neumann algebra $\mathcal{A}(\mathcal{M})$. In particular, from the  the Reeh-Schlieder theorem \cite{Haag92,Witten18,SW1,Summers87b}, it turns out that the vacuum state $\vert 0 \rangle$ is both cyclic and separating for the von Neumann algebra $\mathcal{A}$. Therefore, we can make use of the Tomita-Takesaki modular theory \cite{Bratteli97,Witten18,SW1,Summers87b,Weyl23} and introduce the anti-linear unbounded operator $S$ whose action on the von Neumann algebra $\mathcal{A}(\mathcal{M})$ is defined as
\begin{align} 
	S \; a \vert 0 \rangle = a^{\dagger} \vert 0 \rangle, \qquad \forall a \in \mathcal{A}(\mathcal{M}),
    \label{TT1}
\end{align}  
from which it follows that $S^2 = 1$ and $S \vert 0 \rangle = \vert 0 \rangle$. By performing a polar decomposition of the operator $S$ \cite{Bratteli97,Witten18,SW1,Summers87b,Weyl23}, one gets
\begin{align}
    S = J  \Delta^{1/2},
    \label{PD}    
\end{align} 
where $J$ is anti-unitary  and $\Delta$ is positive and self-adjoint. These modular operators satisfy the following properties \cite{Bratteli97,Witten18,SW1,Summers87b,Weyl23}:
\begin{align} 
	J \Delta^{1/2} J &= \Delta^{-1/2}, \quad \,\,	\Delta^\dagger = \Delta, \nonumber \\
	S^{\dagger} &= J \Delta^{-1/2},  \,\,\,\,\, J^{\dagger} = J, \nonumber \\
	\Delta &= S^{\dagger} S,  \quad \,\,\,\,\,\, J^2 = 1.
    \label{TTP}
\end{align}
According to the Tomita-Takesaki theorem \cite{Bratteli97,Witten18,SW1,Summers87b,Weyl23}, one has that   $J \mathcal{A}(\mathcal{M}) J = \mathcal{A}'(\mathcal{M})$, that is, upon conjugation by the operator $J$, the algebra $\mathcal{A}(\mathcal{M})$ is mapped into its commutant $\mathcal{A'}(\mathcal{M})$, namely: 
\begin{equation} 
\mathcal{A'}(\mathcal{M}) = \{ \; a' \, \vert \; [a,a']=0, \forall a \in \mathcal{A}(\mathcal{M}) \;\}.
    \label{commA}
\end{equation} 
The Tomita-Takesaki modular theory is particularly suited for the analysis of the Bell-CHSH inequality within the framework of relativistic Quantum Field Theory \cite{SW1,Summers87b}. As shown in \cite{Weyl23},  it gives a way of constructing in a purely algebraic way Bob's  operators from Alice's ones by making use of the modular conjugation $J$. That is, given Alice's operator $A_f$, one can assign the operator $B_f = J A_f J$ to Bob, with the guarantee that they commute with each other since by the Tomita-Takesaki theorem the operator $B_f = J A_f J$ belongs to the commutant $\mathcal{A'}(\mathcal{M})$ \cite{Weyl23}. \\\\A very useful result on the Tomita-Takesaki modular theory, proven by  \cite{Rieffel77,Eckmann73}, enables one to lift the action of the modular operatos $(J, \Delta)$ to the space of the test functions. In fact, when equipped with the Lorentz-invariant inner product $\langle f \vert g\rangle$, Eq.\eqref{inner}, the set of test functions give rise to a complex Hilbert space $\mathcal{F}$ which enjoys several features. More precisely,  it turns out that the subspaces $\mathcal{M}$ and $i\mathcal{M}$ are standard subspaces for $\mathcal{F}$ \cite{Rieffel77}, meaning that:  i) $\mathcal{M} \cap i \mathcal{M} = \{ 0 \}$; ii) $\mathcal{M} + i \mathcal{M}$ is dense in $\mathcal{F}$. According to \cite{Rieffel77}, for such subspaces it is possible to set a modular theory analogous to that of the Tomita-Takesaki. One introduces an operator $s$ acting on $\mathcal{M} + i\mathcal{M}$ as
\begin{align}
    s (f+ih) = f-ih. \;, 
    \label{saction}
\end{align}
for $f,h \in \mathcal{M}$. Notice that with this definition, it follows  that $s^2 = 1$. Using the  polar decomposition, one has:  
\begin{align}
    s = j \delta^{1/2},
\end{align}
where $j$ is an anti-unitary operator and $\delta$ is  positive and self-adjoint.  Similarly to the operators $(J,\, \Delta)$, the  operators $(j,\, \delta)$ fulfill  the following properties \cite{Rieffel77}:
\begin{align}
    j \delta^{1/2} j &= \delta^{-1/2}, \,\,\,\,\,\,  \delta^\dagger = \delta\nonumber \\
    s^\dagger &= j \delta^{-1/2}, \,\,\, j^\dagger = j \nonumber \\
    \delta &= s^\dagger s, \,\,\,\,\,\,\,\,\,\,\, j^2=1.
\end{align}
Moreover, as shown in  \cite{Rieffel77},   a test function $f$ belongs to $\mathcal{M}$ if and only if 
\begin{equation} 
s f = f \;. 
    \label{sff}
\end{equation}
In fact, suppose that $f \in \mathcal{M}$. On general grounds, owing to Eq.\eqref{saction}, one writes 
\begin{equation}
sf = h_1 + i h_2 \;, 
    \label{pv1}
\end{equation}
for some $(h_1,h_2)$. Since $s^2=1$ it follows that 
\begin{equation} 
f = s(h_1 + i h_2) = h_1 -i h_2  \;, \label{pv2}
\end{equation} 
so that $h_1=f$ and $h_2=0$. In much the same way, one has that $f' \in \mathcal{M}'$ if and only if $s^{\dagger} f'= f'$. \\\\The lifting of the action of the operators $(J, \Delta)$ to the space of test functions is thus achieved by \cite{Eckmann73} 
\begin{align} 
 J e^{i {\varphi}(f) } J  = e^{-i {\varphi}(jf) }, \quad \Delta e^{i {\varphi}(f) } \Delta^{-1} = e^{i {\varphi}(\delta f) }. \label{jop}
\end{align} 
Also, it is worth noting that if $f \in \mathcal{M} \implies jf \in \mathcal{M}'$. This property follows from 
\begin{equation} 
s^{\dagger} (jf) = j \delta^{-1/2} jf = \delta f = j (j\delta f) = j (sf) = jf \;. \label{jjf} 
\end{equation} 
It is also worth reminding that, in the case of wedge regions in Minkowski spacetime, the spectrum of $\delta$ coincides with the positive real line, {\it i.e.}, $\log(\delta) = \mathbb{R}$ \cite{Bisognano75}, being an unbounded operator with continuous spectrum.\\\\We have now all ingredients for the evaluation of the correlation functions of the Weyl operators. Looking at expression   \eqref{trace rhoABphi over phi delta}, it is easy to realize that te basic quantity to be computed is of the kind
\begin{equation}
\langle e^{i \varphi(f_A)} e^{ \pm i\varphi(f_B)}\rangle = \langle e^{i( \varphi(f_A)\pm \varphi(f_B))}\rangle = e^{-\frac{1}{2} ||f_A \pm f_B||^2} \;, \label{Wex}
\end{equation}
so that  we need to evaluate  the following norms $(||f_A||^2, ||f_B||^2) $ and the inner product $\langle f_A | f_B \rangle$. We focus first on Alice's test function $f_A$. We require that $f_A \in {\cal M(O)}$ where ${\cal O}$ is taken to be located in the right Rindler wedge. Following \cite{SW1,Summers87b,Weyl23}, the test function $f_A$ can be further specified by relying on the spectrum of the operator $\delta$. Ppicking up the spectral subspace specified by $[\lambda^2-\varepsilon, \lambda^2+\varepsilon ] \subset (0,1)$ and introducing  the normalized vector $\phi$ belonging to this subspace, one writes 
\begin{equation}
f _A = \eta  (1+s) \phi \;,
\label{nmf}
\end{equation}
where $\eta$ is an arbitrary parameter. As required by the setup outlined above, equation \eqref{nmf} ensures that 
\begin{equation}
s f_A = f_A  \;. \label{fafa}
\end{equation}
We notice that $j\phi$ is orthogonal to $\phi$, {\it i.e.}, $\langle \phi |  j\phi \rangle = 0$. In fact, from 
\begin{align} 
\delta^{-1} (j \phi) =  j (j \delta^{-1} j) \phi = j (\delta \phi), 
\label{orth}
\end{align}
it follows that the modular conjugation $j$ exchanges the spectral subspace $[\lambda^2-\varepsilon, \lambda^2+\varepsilon ]$ into $[1/\lambda^2-\varepsilon,1/ \lambda^2+\varepsilon ]$.  Concerning now Bob's test function $f_B$, we make use of the modular conjugation operator $j$ and define 
\begin{equation} 
f_B = j f_A \;, \label{fb}
\end{equation}
so that 
\begin{equation} 
s^{\dagger} f_B = f_B
\end{equation}
meaning that, as required by the relativistic causality, $f_B$ belongs to the symplectic  complement $\mathcal{M'}(\mathcal{O})$, located in the left Rindler wedge, namely: $f_B \in  \mathcal{M'}(\mathcal{O})$. Finally, taking into account that $\phi$ belongs to the spectral subspace $[\lambda^2-\varepsilon, \lambda^2+\varepsilon ] $, it follows that \cite{Weyl23}, 
\begin{align}
\vert\vert f_A \vert\vert^2  &= \vert\vert jf _A \vert\vert^2 = \eta^2 (1+\lambda^2) \nonumber \\
\langle f_A \vert jf_A \rangle &= 2 \eta^2 \lambda  \;. \label{sfl}
\end{align}

\section{Analysis of the Bell-CHSH inequality}\label{BCHSH}

\begin{figure}[t!]
	\begin{minipage}[b]{1.0\linewidth}
		\includegraphics[width=\textwidth]{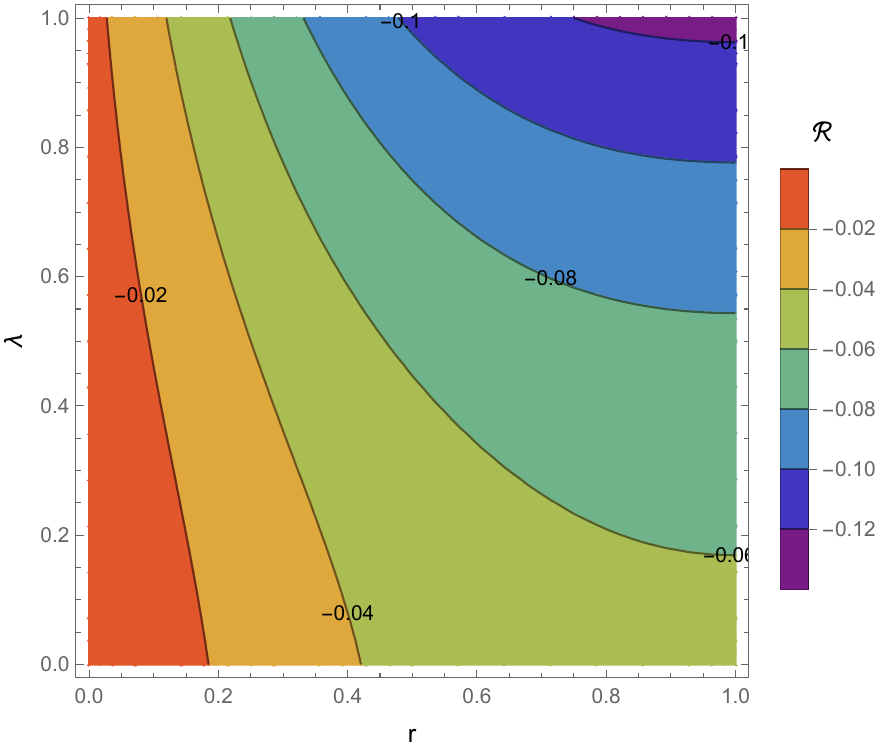}
	\end{minipage} \hfill
\caption{Contour plot exhibiting the negative behavior of $\cal R$ for  as a function of $\lambda$ and $r$, for $\omega_A=0.5$, $\omega_B=0.6$, $\eta=0.1$. The four Bell's  angles $(\alpha,\alpha', \beta, \beta')$ are chosen as in Eq.\eqref{angles}.}
	\label{Fig1}
	\end{figure}

\begin{figure}[t!]
	\begin{minipage}[b]{1.0\linewidth}
		\includegraphics[width=\textwidth]{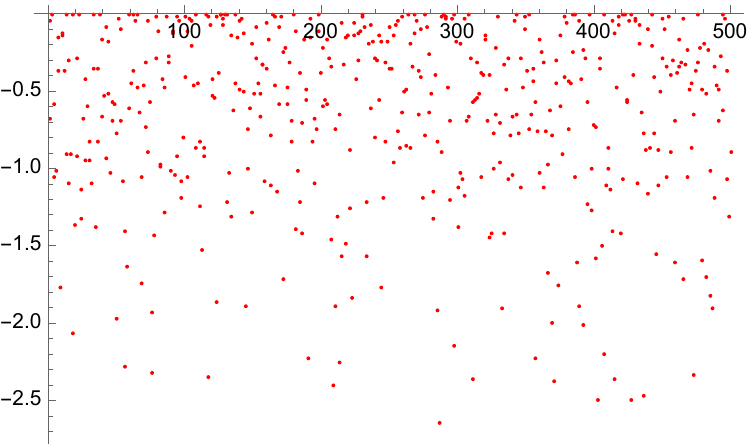}
	\end{minipage} \hfill
\caption{Plot exhibiting the negative behavior of $\cal R$ for randomly chosen values of the parameters $(\lambda, r, \omega_A, \omega_B, \eta)$.  The four Bell's angles $(\alpha,\alpha', \beta, \beta')$ are  chosen as in Eq.\eqref{angles}. The $y$-axis refers to ${\cal R}$ and the $x$-axis is for the samples. }
	\label{Fig2}
	\end{figure}

We are now ready to investigate the Bell-CHSH inequality, Eq.\eqref{ccorr}. Let us 
begin by defining the Bell operators \cite{SW1,Summers87b,Sorella:2023iwz}: 
\begin{eqnarray}
    A \vert g_{A} \rangle &= & e^{i \alpha} \vert e_{A} \rangle \;, \qquad 
    A \vert e_{A} \rangle =  e^{-i \alpha} \vert g_{A} \rangle \nonumber \\
    A' \vert g_{A} \rangle &= & e^{i \alpha'} \vert e_{A} \rangle \;, \qquad 
    A' \vert e_{A} \rangle =  e^{-i \alpha'} \vert g_{A} \rangle \nonumber \\
    B \vert g_{B} \rangle &= & e^{-i \beta} \vert e_{B} \rangle \;, \qquad 
    B \vert e_{B} \rangle =  e^{i \beta} \vert g_{B} \rangle \nonumber \\
    B' \vert g_{B} \rangle &= & e^{-i \beta'} \vert e_{B} \rangle \;, \qquad 
    B' \vert e_{B} \rangle =  e^{i \beta'} \vert g_{B} \rangle \;, \label{bopp}
\end{eqnarray}
which fulfill the whole set of conditions \eqref{Bop}.  The parameters $(\alpha, \alpha',\beta, \beta')$ 
are the four Bell's angles entering the Bell-CHSH inequality. These parameters will be chosen at the 
best convenience. \\\\Reminding that the initial state for $AB$ is 
\begin{align}
    \vert \psi \rangle_{AB} = \frac{\vert g_{A} \rangle \otimes \vert g_{B} \rangle + r \vert e_{A} \rangle \otimes \vert e_{B} \rangle}{\sqrt{1+r^2}}, \; \; \; r \in [0,\, 1] \;, \label{rst}
\end{align}
and making use of 
\begin{equation}
    \mu_{A} \vert g_{A} \rangle = e^{i \Omega_{A} \tau_{A}^{0}} \vert e_{A} \rangle\;, \qquad
    \mu_{A} \vert e_{A} \rangle = e^{-i \Omega_{A} \tau_{A}^{0}} \vert g_{A} \rangle,
\end{equation}
and similar expression for $\mu_{B}$, for the Bell-CHSH correlator $\langle C\rangle $ we get
\begin{align}
    \langle \mathcal{C} \rangle &= \frac{2r}{1+r^2} [\cos(\alpha + \beta) \langle c_{A}^{2}c_{B}^{2} \rangle \nonumber \\
    &\quad + \cos(\alpha - \beta + 2 \omega_B) \langle c_{A}^{2}s_{B}^{2} \rangle \nonumber \\
    &\quad + \cos(\alpha - \beta - 2 \omega_A) \langle s_{A}^{2}c_{B}^{2} \rangle \nonumber \\
    &\quad + \cos(\alpha + \beta - 2 \omega_A - 2 \omega_B) \langle s_{A}^{2}s_{B}^{2} \rangle] \nonumber \\
    &\left. \quad + 2 \cos(\alpha - \omega_{A} - \beta + \omega_{B}) \langle c_{A}s_{A}c_{B}s_{B} \rangle \right. \nonumber \\
    &\quad - 2 \cos(\alpha - \omega_{A} + \beta - \omega_{B}) \langle c_{A}s_{A}c_{B}s_{B} \rangle \nonumber \\
    &\quad+ (\alpha \rightarrow \alpha') + (\beta \rightarrow \beta') \nonumber \\
    &\quad- (\alpha \rightarrow \alpha',\, \beta \rightarrow \beta'),
\end{align}
where $\omega_{A} \equiv \Omega_{A}\tau_{A}^{0}$ and $\omega_{B} \equiv \Omega_{B}\tau_{B}^{0}$.  Furthermore, by employing expressions \eqref{sfl}, it follows that 
\begin{align}
    \langle \mathcal{C} \rangle = &\frac{2r}{1+r^2} \left\{ \frac{1}{4}\cos(\alpha + \beta)\left[1 + 2e^{-2(1+\lambda^2)\eta^2} \right. \right. \nonumber \\
    &+ \frac{1}{2}e^{-4(1+\lambda)^2\eta^2} + \frac{1}{2}e^{-4(1-\lambda)^2\eta^2} \left. \right ] \nonumber \\
    &+ \frac{1}{4}\left(\cos(\alpha - \beta + 2 \omega_B) + \cos(\alpha - \beta + 2 \omega_A)\right) \nonumber \\
    &\times \left[1 - \frac{1}{2}e^{-4(1+\lambda)^2\eta^2} - \frac{1}{2}e^{-4(1-\lambda)^2\eta^2}\right] \nonumber \\
    &+ \frac{1}{4}\cos(\alpha + \beta - 2 \omega_A - 2 \omega_B) \nonumber \\
    &\times \left. \left[1 - 2e^{-2(1+\lambda^2)\eta^2} + \frac{1}{2}e^{-4(1+\lambda)^2\eta^2} + \frac{1}{2}e^{-4(1-\lambda)^2\eta^2} \right] \right\} \nonumber \\
    &+ \left(\cos(\alpha - \beta - \omega_A + \omega_B) - \cos(\alpha + \beta - \omega_A - \omega_B)\right) \nonumber \\
    &\times \frac{1}{4}\left[\frac{1}{2}e^{-4(1-\lambda)^2\eta^2} - \frac{1}{2}e^{-4(1+\lambda)^2\eta^2}\right] \nonumber \\
    &+ (\alpha \rightarrow \alpha') + (\beta \rightarrow \beta') - (\alpha \rightarrow \alpha',\, \beta \rightarrow \beta').  \label{ffb}
\end{align}
From this expression one learns several things: 
\begin{itemize} 
\item the contribution arising from the scalar field $\varphi$ is encoded in the terms containing the exponentials $e^{-4 \eta^2(1\pm\lambda)^2}$ and $e^{-2\eta^2(1+\lambda^2)}$. It is worth reminding here that the parameter $\eta^2$ is related to the norm of the test function $f_A$, Eqs.\eqref{sfl}.
\item when the quantum field $\varphi$ is removed, {\it i.e.} $\eta^2 \rightarrow 0$, expression \eqref{ffb} reduces to the usual Bell-CHSH inequality, namely 
\begin{eqnarray}
\langle {\cal C} \rangle_{\eta=0} & = & \frac{2r}{1+r^2} \left(  \cos(\alpha + \beta) +\cos(\alpha' +\beta) \right) \nonumber \\
& + & \frac{2r}{1+r^2}\left( \cos(\alpha + \beta')- \cos(\alpha' + \beta')\right) \;. \label{cbell}
\end{eqnarray}
It is known that the angular part of Eq.\eqref{cbell} is maximized by \cite{SW1,Summers87b}:
\begin{equation} 
\alpha= 0\;, \qquad \beta =-\frac{\pi}{4}\; \qquad \alpha'= \frac{\pi}{2}\;, \qquad \beta'=\frac{\pi}{4} \;, \label{angles}
\end{equation}
yielding 
\begin{equation}
\langle {\cal C} \rangle_{\eta=0}  = 2\sqrt{2}  \frac{2r}{1+r^2} \;, \label{q1}
\end{equation}
which, for a maximally entangled state, $r=1$, gives Tsirelson's bound 
\begin{equation}
\langle {\cal C} \rangle_{\eta=0,r=1}  = 2\sqrt{2}   \;. \label{q2}
\end{equation}
\item however, when $\eta^2\neq 0$, {\it i.e.} when the quantum field $\varphi$ is present, the exponential factors  $e^{-4 \eta^2(1\pm\lambda)^2}$ and $e^{-2\eta^2(1+\lambda^2)}$ have the effect of producing a damping, resulting in a decreasing of the violation of the Bell-CHSH inequality as compared to the pure Quantum Mechanical case, as it  can be seen from  Fig.\eqref{Fig1} and Fig.\eqref{Fig2}, where the plot of the quantity
\begin{equation} 
{\cal R} = \langle C \rangle -  \langle C \rangle_{\eta=0} \;, \label{RR}
\end{equation}
is depicted. A damping behavior is signaled by ${\cal R}\le 0$.
\end{itemize} 

\section{The dephasing coupling}\label{dephasing}\label{dephasing}

The damping effect due to the scalar field $\varphi$ may be captured in a very simple way by looking at the so-called dephasing coupling 
\cite{Tjoa:2022vnq,Tjoa:2022lel}, whose corresponding unitary evolution operator reads 
\begin{equation}
    \mathcal{U}_{J} = e^{-i \sigma_{j}^{z}\; \otimes \;\varphi(f_{j})} \; \; \; \; \; \;j=A,\,B
    \label{dephasing unit op}
\end{equation}
with $\mathcal{U} = \mathcal{U}_{A} \otimes \mathcal{U}_{B}$. For the evolved wave function, we have
\begin{equation}
    \frac{\left ( e^{i\varphi(f_{A}+f_{B})} \vert g_{A} \rangle \vert g_{B} \rangle + r e^{-i\varphi(f_{A}+f_{B})} \vert e_{A} \rangle \vert e_{B} \rangle\right )}{(1+r^2)^\frac{1}{2}}  \vert 0 \rangle.
    \label{evolved wave function}
\end{equation}
For the density matrix, we have
\begin{align}
    \rho_{AB\varphi} = & \left[ e^{i\varphi(f_{A}+f_{B})} \vert g_{A} \rangle \vert g_{B} \rangle + r e^{-i\varphi(f_{A}+f_{B})} \vert e_{A} \rangle \vert e_{B} \rangle \right] \vert 0 \rangle \nonumber \\
    & \times  \langle 0 \vert \left[ e^{-i\varphi(f_{A}+f_{B})} \langle g_{A} \vert \langle g_{B} \vert + r e^{i\varphi(f_{A}+f_{B})} \langle e_{A} \vert \langle e_{B} \vert \right].
    \label{density matrix ABphi}
\end{align}
Tracing over $\varphi$
\begin{align}
    \hat{\rho}_{AB} = \frac{1}{1+r^2} & \left[ \vert g_{A} \rangle \vert g_{B} \rangle \langle g_{A} \vert \langle g_{B} \vert \right. \nonumber \\
    & + r e^{-2{\lVert f_{A}+f_{B}\rVert}^2} \vert g_{A} \rangle \vert g_{B} \rangle \langle e_{A} \vert \langle e_{B} \vert \nonumber \\
    & + re^{-2{\lVert f_{A}+f_{B}\rVert}^2} \vert e_{A} \rangle \vert e_{B} \rangle \langle g_{A} \vert \langle g_{B} \vert \nonumber \\
    & \left. + r^2 \vert e_{A} \rangle \vert e_{B} \rangle \langle e_{A} \vert \langle e_{B} \vert \right].
    \label{density matrix AB}
\end{align}
Proceeding as in the previous section, for the Bell-CHSH inequality we get 
\begin{align}
    \langle \mathcal{C} \rangle &= \frac{2r}{1+r^2} e^{-4 \eta^2 (1+\lambda)^2} \left[ \cos(\alpha + \beta) + \cos(\alpha' + \beta) \right. \nonumber\\
    & \left.  + \cos(\alpha + \beta') - \cos(\alpha' + \beta') \right] \;,
    \label{chsh dephasing}
\end{align}
which clearly exhibits a decreasing with respect to the case in which the field is absent.

\section{Further considerations} \label{fc} 

\begin{figure}[t!]
	\begin{minipage}[b]{1.0\linewidth}
		\includegraphics[width=\textwidth]{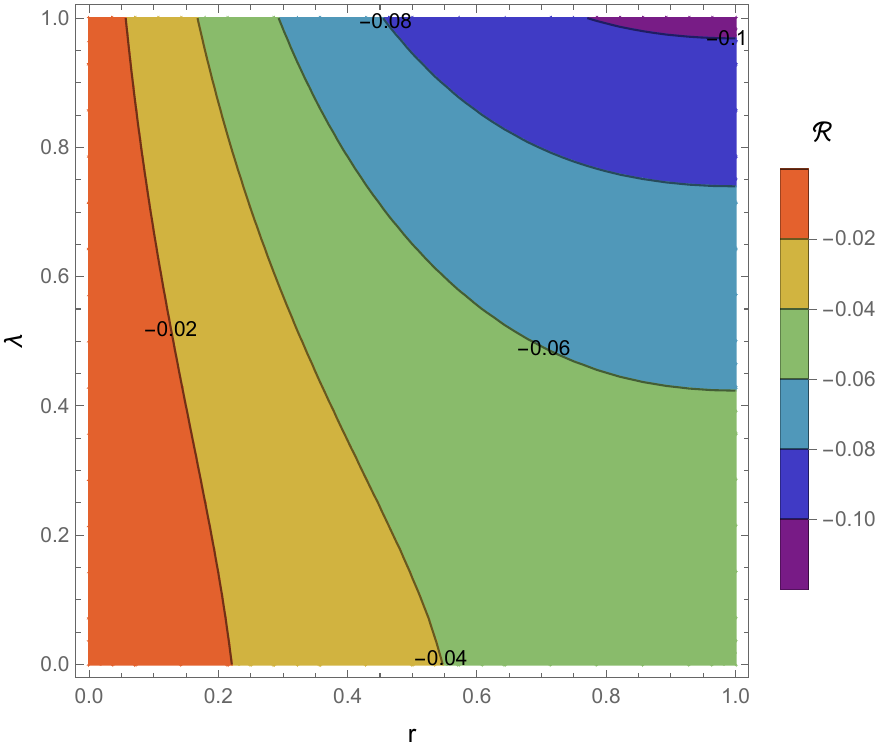}
	\end{minipage} \hfill
\caption{Contour plot exhibiting the negative behavior of $\cal R$   as a function of $\lambda$ and $r$, for $\omega_A=0.5$, $\omega_B=0.6$, 
$\eta=0.1$. The four Bell's  angles $(\alpha,\alpha', \beta, \beta')$ are chosen now as: 
$(\alpha=0, \alpha'=\frac{\pi}{2}, \beta= -\frac{\pi}{6}, \beta'=\frac{\pi}{6})$. }
	\label{Fig3}
	\end{figure}

The relatively simple expression obtained in the case of the dephasing coupling enables us to elaborate more on a few points, providing a better illustration of our findings:

\begin{itemize} 

\item We observe that the angular part of Eq.\eqref{chsh dephasing}, {\it i.e.}
\begin{equation} 
\left( \cos(\alpha + \beta) + \cos(\alpha' + \beta)   + \cos(\alpha + \beta') - \cos(\alpha' + \beta') \right)  \label{n1} 
\end{equation} 
factorizes from the rest of the expression. This implies that the usual choice of the Bell angles given in eq.\eqref{angles} is not mandatory. Any other choice of $(\alpha, \alpha', \beta, \beta' )$ yielding to 
\begin{equation} 
\vert \cos(\alpha + \beta) + \cos(\alpha' + \beta)   + \cos(\alpha + \beta') - \cos(\alpha' + \beta') \vert > 2 \label{n2}
\end{equation} 
leads to the same conclusion: a decreasing of the violation of the Bell-CHSH inequality. 

\item A second point worth to highlight concerns the use of the test function of the form given in Eqs.\eqref{nmf}, \eqref{fb}, namely  
\begin{equation}
f _A = \eta  (1+s) \phi \;, \qquad s f_A = f_A \;, \qquad f_B = j f_A
\label{n3}
\end{equation}
As discussed in  \cite{SW1,Summers87b}, this specific form is dictated by the possibility of taking full profit of the powerful results related to von Neumann algebras and to the Tomita-Takesaki theory. Though, also here, one is not obliged to make this choice. To grasp this point, we go back to the general expression  \eqref{density matrix ABphi}, valid for a generic choice of Alice's and Bob's test functions $(f_A,f_B)$, not subjects to condition \eqref{n3}. For the Bell-CHSH inequality, we would get 
\begin{align}
    \langle \mathcal{C} \rangle &= \frac{2r}{1+r^2} e^{-2 ||f_A + f_B||^2} \left[ \cos(\alpha + \beta) + \cos(\alpha' + \beta) \right. \nonumber\\
    & \left.  + \cos(\alpha + \beta') - \cos(\alpha' + \beta') \right] \;,
    \label{n4}
\end{align}
At this stage, we could leave the test functions $f_A$ and $f_B$ unspecified. In this case, we would not be able to express the norm $||f_A + f_B||^2$ in terms of the parameters $(\eta, \lambda)$. Though, as far as the decreasing of the violation of the Bell-CHSH inequality is concerned, our conclusion remains unaltered. 

\item \ A third aspect is related to the presence of the parameters $(\eta, \lambda)$ in expression \eqref{chsh dephasing}. As already mentioned, the parameter $\lambda$ is related to the spectrum of the modular operator $\delta$ \cite{SW1,Summers87b}. In general, the characterization of the spectrum of $\delta$ is a quite difficult task, being known only in some specific situations as, for instance, in the case in which the spacetime regions considered for Alice and Bob are causal wedges, as the left and right Rindler wedges. In this case, from the analysis of 
Bisognano and Wichmann \cite{Bisognano75}, one learns that $\lambda \in [0, \infty]$, {\it i.e} the modular operator $\delta$ has a continuous spectrum coinciding with the positive real line. One sees thus that the parameter $\lambda$ has a deep meaning: it is directly connected to Alice's and Bob's causal wedge regions. \\\\As for the parameter $\eta$, it reflects the freedom one has in defining the test function $f_A$ through the operator $s$. One notices that equation \eqref{n3} does not fix completely $f_A$. It turns out that $f_A$ is determined up to the value of its norm
, namely 
\begin{equation} 
||f_A||^2 = \langle f_A |  f_A \rangle = \eta^2 (1 +\lambda^2) \label{n5} 
\end{equation}
which is encoded precisely in the parameter $\eta$. As discussed in \cite{Weyl23,Guimaraes:2024alk}, this parameter is a free parameter appearing in the Quantum Field Theory formulation of the Bell-CHSH inequality in terms of Weyl operators $W_{f_A} = e^{i {\varphi}(f_A) }$. Needless to say, the operator $W_{f_A}$ remains bounded and unitary for any value of the parameter $\eta$. In other words, $\eta$ is akin to the free Bell's angles $(\alpha, \alpha', \beta, \beta')$ and can be chosen at the best convenience, see \cite{Weyl23,Guimaraes:2024alk}.

\item The previous remark applies to the case of the $\delta$-coupled detectors as well. Let us illustrate this point by adding a third plot, Fig.\eqref{Fig3}, in which the negativity of the factor $\cal R$ is displayed for a different value of the Bell's angles $(\alpha, \alpha', \beta, \beta')$: 
\begin{equation} 
\alpha=0 \; \qquad \alpha'=\frac{\pi}{2} \;, \qquad  \beta= -\frac{\pi}{6} \;, \qquad \beta'=\frac{\pi}{6} \;. \label{nang}
\end{equation} 
These angles yield the non-maximal value 2.732 for the angular part \eqref{n2}. The new plot exhibits the same pattern displayed by Fig.\eqref{Fig2}

\end{itemize}

\section{Conclusions}
In this work we have analyzed the interaction between a spin $1/2$ Unruh-De Witt detector and a relativistic quantum scalar field $\varphi$. Emphasis has been placed on a thorough examination of the effects arising from the presence of the scalar field on the Bell-CHSH inequality. \\\\In particular, in the cases involving the so-called $\delta$-coupled detector and the dephasing channel,  we evaluated the  influence of the scalar field in closed form. That was possible due to the use of the von Neumann algebra of the Weyl operators and of the  powerful Tomita-Takesaki modular theory, especially well-suited for the study of the Bell-CHSH inequality in Quantum Field Theory. \\\\The main result of the present investigation  is that the presence of a scalar quantum field theory causes a damping effect, resulting in a decreasing of the violation of the Bell-CHSH inequality as compared to the case in which the field is absent. \\\\To some extent, this behavior can be traced back to the fact that, in the case of spin $1/2$, the pure Quantum Mechanical Bell-CHSH inequality attains Tsireslon's bound, $2 \sqrt{2}$, which is the maximum allowed value. As such, one could expect that the presence of a quantum scalar field can give rise to a decreasing of the value of the violation, as reported in Figs. \eqref{Fig1} and \eqref{Fig2}. \\\\As a future investigation, we are already considering the case of the interaction between a spin 1 detector, {\it i.e.} a pair q-trits, and a scalar field. This system is of particularly interest due to the well known feature that, for a spin 1, the Tsirelson bound is not achieved in Quantum Mechanics, see \cite{Gisin,Peruzzo:2023nrr}. Rather, the maximum value attained is $\frac{2}{3}(1+ 2\sqrt{2}) \sim 2.5$. One sees thus that, in the case of spin 1, there is a small allowed window, namely $[\frac{2}{3}(1+ 2\sqrt{2}), 2\sqrt{2}]$, which, unlike the case of spin $1/2$, might yield to a potential increase of the value of the violation of the Bell-CHSH inequality due to the interaction with a scalar quantum field \cite{prep}.

\section*{Acknowledgments}
	The authors would like to thank the Brazilian agencies CNPq and CAPES, for financial support.  S.P.~Sorella, I.~Roditi, and M.S.~Guimaraes are CNPq researchers under contracts 301030/2019-7, 311876/2021-8, and 310049/2020-2, respectively.

	
\end{document}